\documentclass[aps,onecolumn,showpacs,showkeys,nofootinbib]{revtex4}
\usepackage{epsfig}
\usepackage{amsmath}
\usepackage{amsfonts}
\usepackage{amssymb}
\usepackage{graphicx}
\usepackage{colordvi}
\begin{document}

\title{The Post-Newtonian Limit of $f(R)$-gravity in the Harmonic Gauge}

\author{A. Stabile\footnote{e -
mail address: arturo.stabile@gmail.com}}

\affiliation{Dipartimento di Ingegneria, Universita'
del Sannio\\Corso Garibaldi, I - 80125 Benevento, Italy}

\begin{abstract}

A general analytic procedure is developed for the post-Newtonian
limit of $f(R)$-gravity with metric approach in the Jordan frame
by using the harmonic gauge condition. In a pure perturbative
framework and by using the Green function method a general scheme
of solutions up to $(v/c)^4$ order is shown. Considering the
Taylor expansion of a generic function $f$ it is possible to
parameterize the solutions by derivatives of $f$. At Newtonian
order, $(v/c)^2$, all more important topics about the Gauss and
Birkhoff theorem are discussed. The corrections to "standard"
gravitational potential ($tt$-component of metric tensor)
generated by an extended uniform mass ball-like source are
calculated up to $(v/c)^4$ order. The corrections, Yukawa and
oscillating-like, are found inside and outside the mass
distribution. At last when the limit $f\rightarrow R$ is
considered the $f(R)$-gravity converges in General Relativity at
level of Lagrangian, field equations and their solutions.

\end{abstract}
\pacs{04.25.Nx; 04.50.Kd; 04.40.Nr}
\keywords{Alternative theories of gravity; newtonian and post-newtonian limit; weak field limit.}
\maketitle

\section{Introduction}

The study of possible modifications of Einstein's theory of
gravity has a long history which reaches back to the early 1920s
\cite{Weyl:1918,Pauli:1919,Bach:1921}. Corrections to the
gravitational Lagrangian, leading to higher-order field equations,
were already studied by several authors
\cite{Weyl:1921,Eddington:1924,Lanczos:1931} shortly after General
Relativity (GR) was proposed. Developments in the 1960s and 1970s
\cite{Buchdahl:1962,DeWitt:1965,Bicknell:1974,Havas:1977,Stelle:1978},
partly motivated by the quantization schemes proposed at that
time, made clear that theories containing {\it only} a $R^2$ term
in the Lagrangian were not viable with respect to their weak field
behavior. Buchdahl, in 1962 \cite{Buchdahl:1962}, rejected pure
$R^2$ theories because of the non-existence of asymptotically flat
solutions.

In  recent years, the effort to give a  physical explanation to
the today observed cosmic acceleration \cite{sneIa,lss,cmbr} has
attracted a good amount of interest in $f(R)$-gravity, considered
as a viable mechanism to explain the cosmic acceleration by
extending the geometric sector of field equations without the
introduction of dark matter and dark energy. There are several
physical and mathematical motivations to enlarge GR by these
theories. For comprehensive review, see
\cite{GRGreview,OdintsovLadek,farhoudi}.

Other issues as, for example, the observed Pioneer anomaly problem
\cite{anderson} can be framed into the same approach
\cite{bertolami} and then, apart the cosmological dynamics, a
systematic analysis of such theories urges at short scale and in
the low energy limit.

While it is very natural to extend Einstein's gravity to theories
with additional geometric degrees of freedom, recent attempts
focused on the old idea of modifying the gravitational Lagrangian
in a purely metric framework, leading to higher-order field
equations. Due to the increased complexity of the field equations
in this framework, the main body of works dealt with some formally
equivalent theories, in which a reduction of the order of the
field equations was achieved by considering the metric and the
connection as independent objects \cite{Francaviglia}.

In addition, many authors exploited the formal relationship to
scalar-tensor theories to make some statements about the weak
field regime \cite{olmo}, which was already worked out for
scalar-tensor theories \cite{Damour:Esposito-Farese:1992}. Also a
Post-Newtonian parameterization with metric approach in the Jordan
Frame has been considered \cite{clifton}.

In this paper, we study the Post Newtonian limit of $f(R)$ in the
harmonic gauge. We are going to focus on the small velocity and
weak field limit within the metric approach. In principle, any
alternative or extended theory of gravity should allow to recover
positive results of General Relativity. It will be very important
to check at any level of our modified theory we can cover the
outcomes of GR.

The plan of the paper is the following. In the
Sec.\ref{fieldequations}, we report the complete scheme of
Newtonian and post-Newtonian limit of field equations for
$f(R)$-gravity and their formal solutions in the harmonic gauge
condition. General comments about the mathematical properties of
equations, their relative solutions (Gauss and Birkhoff theorem)
and Minkowskian behavior of metric tensor are reported. In the
Sec. \ref{ball_like} we show the complete solutions (Newtonian and
post-Newtonian level) when an uniform mass ball-like source is
considered. The point-like source limit of the newtonian solution
is considered and the compatibility of $f(R)$-gravity with respect
to GR is shown.
Concluding remarks are drawn in Sec. \ref{conclusions}.

\section{The Field equations up to post-Newtonian level}\label{fieldequations}

Let us start with a general class of higher order theories given
by the action

\begin{eqnarray}\label{HOGaction}\mathcal{A}=\int d^{4}x\sqrt{-g}[f(R)+\mathcal{X}\mathcal{L}_m]
\end{eqnarray}
where $f$ is an unspecified function of curvature invariant $R$.
The term $\mathcal{L}_m$ is the minimally coupled ordinary matter
contribution. In the metric approach, the field equations are
obtained by varying (\ref{HOGaction}) with respect to
$g_{\mu\nu}$. We get

\begin{eqnarray}\label{fieldequationHOG}
H_{\mu\nu}\,=\,f'R_{\mu\nu}-\frac{f}{2}g_{\mu\nu}-f_{;\mu\nu}+g_{\mu\nu}\Box
f'=\mathcal{X}\,T_{\mu\nu}
\end{eqnarray}

\begin{eqnarray}\label{tracefieldequationHOG}
H\,=\,g^{\alpha\beta}H_{\alpha\beta}\,=\,3\Box
f'+f'R-2f\,=\,\mathcal{X}\,T
\end{eqnarray}
Here,
$T_{\mu\nu}=\frac{-2}{\sqrt{-g}}\frac{\delta(\sqrt{-g}\mathcal{L}_m)}{\delta
g^{\mu\nu}}$ is the the energy-momentum tensor of matter, while
$T=T^{\sigma}_{\,\,\,\,\,\sigma}$ is the trace,
$f'=\frac{df(R)}{dR}$, $\Box={{}_{;\sigma}}^{;\sigma}$ and
$\mathcal{X}\,=\,8\pi G$\footnote{Here we use the convention
$c\,=\,1$.}. The conventions for Ricci's tensor is
$R_{\mu\nu}={R^\sigma}_{\mu\sigma\nu}$ while for the Rienman
tensor is
${R^\alpha}_{\beta\mu\nu}=\Gamma^\alpha_{\beta\nu,\mu}+...$. The
affinities are the usual Christoffel's symbols of the metric:
$\Gamma^\mu_{\alpha\beta}=\frac{1}{2}g^{\mu\sigma}(g_{\alpha\sigma,\beta}+g_{\beta\sigma,\alpha}
-g_{\alpha\beta,\sigma})$. The adopted signature is $(+---)$ (see
for the details \cite{landau}).

The paradigm of post-Newtonian limit is starting from a develop of
metric tensor (and of all additional fields in the theory) with
respect to dimensionless quantity $v$. A system of moving bodies
radiates gravitational waves and thus loses energy. This loss
appears only in the fifth approximation in $v$. In the first four
approximations, the energy of the system remains constant. From
this it follows that a system of gravitating bodies can be
described by a Lagrangian correctly to terms of order $v^4$ in the
absence of an electromagnetic field, for which a Lagrangian exists
in general only to terms of second order. We thus find the
equations of motion of the system in the next approximation after
the Newtonian.

To solve our problem we must start with the determination, in this
same approximation of the weak gravitational field, of the metric
tensor $g_{\mu\nu}$ (for details see \cite{newtonian_limit_fR})

\begin{eqnarray}\label{metric_tensor_PPN}
  g_{\mu\nu}\,\sim\,\begin{pmatrix}
  1+g^{(2)}_{tt}(t,\mathbf{x})+g^{(4)}_{tt}(t,\mathbf{x})+\dots & g^{(3)}_{ti}(t,\mathbf{x})+\dots \\
  g^{(3)}_{ti}(t,\mathbf{x})+\dots & -\delta_{ij}+g^{(2)}_{ij}(t,\mathbf{x})+\dots\end{pmatrix}
\end{eqnarray}
The set of coordinates\footnote{The greek index runs between $0$
and $3$; the latin index between $1$ and $3$.} adopted is
$x^\mu\,=\,(t,x^1,x^2,x^3)$. The Ricci scalar becomes

\begin{eqnarray}
R\,\sim\,R^{(2)}(t,\mathbf{x})+R^{(4)}(t,\mathbf{x})+\dots
\end{eqnarray}
The $n$-th derivative of Ricci function can be developed as

\begin{eqnarray}
f^{n}(R)\,\sim\,f^{n}(R^{(2)}+R^{(4)}+\dots)\,\sim\,f^{n}(0)+f^{n+1}(0)R^{(2)}+f^{n+1}(0)R^{(4)}+\frac{1}{2}f^{n+2}(0)
{R^{(2)}}^2+\dots
\end{eqnarray}

From lowest order of field equations (\ref{fieldequationHOG}) we
have

\begin{eqnarray}\label{PPN-field-equation-general-theory-fR-O0}
f(0)\,=\,0
\end{eqnarray}
which trivially follows from the above assumption
(\ref{metric_tensor_PPN}) that the space-time is asymptotically
Minkowskian. This result suggests a first  consideration. If the
Lagrangian is developable around a vanishing value of the Ricci
scalar the relation
(\ref{PPN-field-equation-general-theory-fR-O0}) will imply that
the cosmological constant contribution has to be zero whatever is
the $f(R)$-gravity theory. This result appears quite obvious but
sometime it is not considered in literature
\cite{spher_symm_fR}.

The Eqs. (\ref{fieldequationHOG}) and
(\ref{tracefieldequationHOG}) at $\mathcal{O}(2)$ - order
(Newtonian level) become

\begin{eqnarray}\label{PPN-field-equation-general-theory-fR-O2}
\left\{\begin{array}{ll}
H^{(2)}_{tt}=f'(0)R^{(2)}_{tt}-\frac{f'(0)}{2}R^{(2)}-f''(0)\triangle
R^{(2)}=\mathcal{X}\,T^{(0)}_{tt}\\\\
H^{(2)}_{ij}=f'(0)R^{(2)}_{ij}+\biggl[\frac{f'(0)}{2}R^{(2)}+f''(0)\triangle
R^{(2)}\biggr]\delta_{ij}-f''(0)R^{(2)}_{,ij}=0\\\\
H^{(2)}=-3f''(0)\triangle
R^{(2)}-f'(0)R^{(2)}=\mathcal{X}\,T^{(0)}
\end{array}\right.
\end{eqnarray}
where $\triangle$ is the Laplacian in the flat space, while at
$\mathcal{O}(3)$ - order become

\begin{eqnarray}\label{PPN-field-equation-general-theory-fR-O3}
H^{(3)}_{ti}\,=\,f'(0)R^{(3)}_{ti}-f''(0){R^{(2)}}_{,ti}\,=\,\mathcal{X}\,T^{(1)}_{ti}
\end{eqnarray}

The solution for the gravitational potential $g^{(2)}_{tt}/2$ has
a Yukawa-like behavior (\cite{newtonian_limit_fR}) depending by a
characteristic length on which it evolves. Besides the Birkhoff
theorem at Newtonian level is modified: the solution can be only
factorized with a function space-depending and an arbitrary
function time depending (\cite{newtonian_limit_fR}). Still more
the corrections to the gravitational potential and the
gravito-magnetic effects
(\ref{PPN-field-equation-general-theory-fR-O3}) are depending on
the only first two derivatives of $f$ in $R\,=\,0$. So different
theories from the third derivative admit the same newtonian
solution.

Remembering the expressions of Christoffel symbols and using the
following approximation for the determinant of metric tensor
$\ln\sqrt{-g}\sim\frac{1}{2}[g^{(2)}_{tt}-g^{(2)}_{mm}]+\dots$, at
$\mathcal{O}(4)$ - order we have,

\begin{eqnarray}\label{PPN-field-equation-general-theory-fR-O4}
\left\{\begin{array}{ll}
H^{(4)}_{tt}\,=\,f'(0)R^{(4)}_{tt}+f''(0)R^{(2)}R^{(2)}_{tt}-\frac{f'(0)}{2}R^{(4)}-\frac{f'(0)}{2}g^{(2)}_{tt}R^{(2)}-
\frac{f''(0)}{4}
{R^{(2)}}^2\\\\\,\,\,\,\,\,\,\,\,\,\,\,\,\,\,\,\,\,\,\,\,-f''(0)\biggl[g^{(2)}_{mn,m}{R^{(2)}}_{,n}+\triangle
R^{(4)}+g^{(2)}_{tt}\triangle R^{(2)}
+g^{(2)}_{mn}{R^{(2)}}_{,mn}-\frac{1}{2}\nabla
g^{(2)}_{mm}\cdot\nabla R^{(2)}\biggr]\\\\\,\,\,\,\,\,\,\,\,\,\,\,
\,\,\,\,\,\,\,\,\,-f'''(0)\biggl[|\nabla
R^{(2)}|^2+R^{(2)}\triangle
R^{(2)}\biggr]=\mathcal{X}\,T^{(2)}_{tt}\\\\\\
H^{(4)}=-3f''(0)\triangle
R^{(4)}-f'(0)R^{(4)}-3f'''(0)\biggl[|\nabla
R^{(2)}|^2+R^{(2)}\triangle
R^{(2)}\biggr]\\\\\,\,\,\,\,\,\,\,\,\,\,\,\,\,\,\,\,\,\,\,\,+3f''(0)\biggl[R^{(2)}_{,tt}-g^{(2)}_{mn}R^{(2)}_{,mn}-
\frac{1}{2}\nabla(g^{(2)}_{tt}-g^{(2)}_{mm})\cdot\nabla
R^{(2)}-g^{(2)}_{mn,m}R^{(2)}_{,n}\biggr]=\mathcal{X}\,T^{(2)}
\end{array}\right.
\end{eqnarray}
where $\nabla$ is the gradient in the flat space. Note that the
propagation of Ricci scalar $R^{(4)}$ has the same dynamics of
previous one. The complete knowledge of correction at fourth order
for the $tt$-component of Ricci tensor fix the third derivative of
$f$ in $R\,=\,0$. Also at this level there is a degeneracy of
$f(R)$-theory: different theories for only the first three
derivatives admit the same gravitational field without obviously
radiation emission.

To complete the perturbative scheme and later find the solutions
it needs to calculate the Ricci tensor components in
(\ref{PPN-field-equation-general-theory-fR-O2}),
(\ref{PPN-field-equation-general-theory-fR-O3}),
(\ref{PPN-field-equation-general-theory-fR-O4}). After some
calculus (\cite{newtonian_limit_fR}, \cite{phd}) one obtains

\begin{eqnarray}\label{PPN-ricci-tensor}
\left\{\begin{array}{ll}R^{(2)}_{tt}=\frac{1}{2}g^{(2)}_{tt,mm}\\\\R^{(4)}_{tt}=\frac{1}{2}g^{(4)}_{tt,mm}+\frac{1}{2}g^{(2)}
_{mn,m}g^{(2)}_{tt,n}+\frac{1}{2}g^{(2)}_{mn}g^{(2)}_{tt,mn}+\frac{1}{2}g^{(2)}_{mm,tt}-
\frac{1}{4}g^{(2)}_{tt,m}g^{(2)}_{tt,m}-\frac{1}{4}g^{(2)}_{mm,n}g^{(2)}_{tt,n}-g^{(3)}_{tm,tm}\\\\
R^{(3)}_{ti}=\frac{1}{2}g^{(3)}_{ti,mm}-\frac{1}{2}g^{(2)}_{im,mt}-\frac{1}{2}g^{(3)}_{mt,mi}+\frac{1}{2}g^{(2)}_
{mm,ti}\\\\
R^{(2)}_{ij}=\frac{1}{2}g^{(2)}_{ij,mm}-\frac{1}{2}g^{(2)}_{im,mj}-\frac{1}{2}g^{(2)}_{jm,mi}-\frac{1}{2}g^{(2)}_
{tt,ij}+\frac{1}{2}g^{(2)}_{mm,ij}
\end{array}\right.
\end{eqnarray}
which represent the most general expressions without assuming any gauge condition.

\subsection{The Newtonian limit of $f(R)$-gravity}

We want to rewrite and generalize the outcome of (\cite{newtonian_limit_fR}) by introducing the Green function
method (we remember that the Newtonian limit corresponds also to linearization of field equations). Let us start
from the trace equation. The solution for the Ricci scalar $R^{(2)}$ in the third
line of (\ref{PPN-field-equation-general-theory-fR-O2}) is

\begin{eqnarray}\label{scalar_ricci_sol_gen}
R^{(2)}(t,\textbf{x})= \frac{m^2\mathcal{X}}{f'(0)}\int
d^3\mathbf{x}'\mathcal{G}(\mathbf{x},\mathbf{x}')T^{(0)}(t,\mathbf{x}')
\end{eqnarray}
where $m^2\,\doteq\,-\frac{f'(0)}{3f''(0)}$ and
$\mathcal{G}(\mathbf{x},\mathbf{x}')$ is the Green function of
field operator $\triangle-m^2$.

The solution for $g^{(2)}_{tt}$, from the first line of
(\ref{PPN-field-equation-general-theory-fR-O2}) by considering
that $R^{(2)}_{tt}=\frac{1}{2}\triangle g^{(2)}_{tt}$, is

\begin{eqnarray}\label{new_sol}
g^{(2)}_{tt}(t,\mathbf{x})=-\frac{\mathcal{X}}{2\pi f'(0)}\int
d^3\textbf{x}'\frac{T^{(0)}_{tt}(t,\textbf{x}')}{|\textbf{x}-
\textbf{x}'|}-\frac{1}{4\pi}\int
d^3\textbf{x}'\frac{R^{(2)}(t,\textbf{x}')}{|\textbf{x}-
\textbf{x}'|}-\frac{2}{3m^2}R^{(2)}(t,\textbf{x})\end{eqnarray} We
can check immediately that when $f\rightarrow R$ we find
$g^{(2)}_{tt}(t,\textbf{x})\rightarrow-2G\int
d^3\textbf{x}'\frac{\rho(\textbf{x}')}{|\textbf{x}-
\textbf{x}'|}$. The expression (\ref{new_sol}) is the "modified"
gravitational potential (here we have a factor 2) for
$f(R)$-gravity. \emph{A such solution which is the newtonian limit
of $f(R)$-gravity is also gauge-free}.

Since we have a linearized version of field equations a such limit
corresponds to one of Einstein equation and the linear
superposition is satisfied. So the $tt$-component of
energy-momentum tensor is, in this limit, the sum of mass energy
volume density of sources:
$T^{(0)}_{tt}\,=\,\Sigma_aM_a\delta(\mathbf{x}-\mathbf{x}_a)$
where $\delta(\mathbf{x})$ is the delta function.

As it is evident the Gauss theorem is not valid since the force
law is not $\propto|\mathbf{x}|^{-2}$. The equivalence between a
spherically symmetric distribution and point-like distribution is
not valid and how the matter is distributed in the space is very
important (\cite{newtonian_limit_R_Ric}).

In this limit the geodesic equation is the Lagrangian of material
point-like embedded in the gravitational field and its dynamics
follows the Newtonian law:

\begin{eqnarray}
\frac{d^2\mathbf{x}}{dt^2}\,=\,\frac{1}{2}\nabla
g^{(2)}_{tt}(t,\mathbf{x})
\end{eqnarray}

\subsection{The post-Newtonian limit of $f(R)$-gravity in the harmonic gauge}

To simplify the expressions of the components of the Ricci tensor (\ref{PPN-ricci-tensor}) we can use
the condition so-called of \emph{harmonic gauge}: $g^{\rho\sigma}\Gamma^{\mu}_{\rho\sigma}\,=\,0$ (see
\cite{newtonian_limit_fR}). The Ricci tensor assumes the following simpler form:

\begin{eqnarray}\label{PPN-ricci-tensor-HG}
\left\{\begin{array}{ll}R^{(2)}_{ij}=\frac{1}{2}\triangle
g^{(2)}_{ij}\\\\R^{(3)}_{ti}=\frac{1}{2} \triangle
g^{(3)}_{ti}\\\\R^{(4)}_{tt}=\frac{1}{2}\triangle
g^{(4)}_{tt}+\frac{1}{2}g^{(2)}_{mn}g^{(2)}_{tt,mn}-\frac{1}{2}g^{(2)}_{tt,tt}-\frac{1}{2}|\bigtriangledown
g^{(2)}_{tt}|^2\end{array}\right.
\end{eqnarray}

From the field equation
(\ref{PPN-field-equation-general-theory-fR-O3}) we find the
general solution for $g^{(3)}_{ti}$

\begin{eqnarray}\label{postnew_sol_ti}
g^{(3)}_{ti}(t,\textbf{x})=-\frac{\mathcal{X}}{2\pi f'(0)}\int
d^3\textbf{x}'\frac{T^{(1)}_{ti}(t,\textbf{x}')}{|\textbf{x}-
\textbf{x}'|}+\frac{1}{6\pi m^2}\frac{\partial}{\partial t}\int
d^3\textbf{x}'\frac{\nabla_{i'}R^{(2)}(t,\textbf{x}')}{|\textbf{x}-
\textbf{x}'|}
\end{eqnarray}
The choice of harmonic gauge enable us to solve with facility the
equation (\ref{PPN-field-equation-general-theory-fR-O3}) but we
lose potential information about the temporal dynamics of
$g^{(2)}_{tt}(t,\textbf{x})$. A such knowledge is very important
to obtain at least in perturbative approach some information about
the Birkhoff theorem. By hypothesizing a perturbative approach
(newtonian-like) we relegated inevitably eventual temporal
dynamics only on the temporal variation of matter source. In fact
in a such hypothesis of work the motion of bodies embedded in
gravitational fields develops very slow with respect to motion of
matter. Then we have ever an instantaneous readjustment of
spacetime. In other words the motion od bodies is adiabatic and it
enables us to factorize the solution and with a time
transformation we get a static solution.

From second line of
(\ref{PPN-field-equation-general-theory-fR-O2}) the solution for
$g^{(2)}_{ij}$ follows

\begin{eqnarray}\label{post_new_ij}
g^{(2)}_{ij}(t,\textbf{x})\,=&\biggl[&\frac{1}{4\pi}\int
d^3\textbf{x}'\frac{R^{(2)}(t,\textbf{x}')}{|\textbf{x}-
\textbf{x}'|}+\frac{2}{3m^2}R^{(2)}(t,\textbf{x})-\frac{1}{6\pi
m^2}\frac{1}{|\mathbf{x}|^3}\int_{\Omega_{|\mathbf{x}|}}d^3\mathbf{x}'R^{(2)}
(t,\mathbf{x}')\biggr]\delta_{ij}\nonumber\\\nonumber\\&+&\biggl[\frac{1}{2\pi
m^2
|\mathbf{x}|^3}\int_{\Omega_{|\mathbf{x}|}}d^3\mathbf{x}'R^{(2)}
(t,\mathbf{x}')-\frac{2}{3m^2}R^{(2)}(t,\mathbf{x})\biggr]\frac{x_i x_j}{|\mathbf{x}|^2}
\end{eqnarray}
where $\Omega_{|\mathbf{x}|}$ represents the integration volume
with radius $|\mathbf{x}|$ (for the details see \cite{weinberg}).
By the solutions (\ref{new_sol}), (\ref{postnew_sol_ti}),
(\ref{post_new_ij}) we can affirm that it is possible to have
solution non-Ricci-flat in vacuum: \emph{Higher Order Gravity
mimics a matter source}. It is evident from (\ref{new_sol}) the
Ricci scalar is a "matter source" which can curve the spacetime
also in absence of ordinary matter. Then it is clear also that the
knowledge of behavior of Ricci scalar inside mass distribution is
fundamental to obtain the behavior of metric tensor outside the
mass.

From the fourth order of field equation, we note also the Ricci
scalar ($R^{(4)}$) propagates with the same $m$ (the second line
of (\ref{PPN-field-equation-general-theory-fR-O4})) and the
solution at second order originates a supplementary matter source
in r.h.s. of (\ref{fieldequationHOG}). The solution is

\begin{eqnarray}\label{Ricci_quarto}
R^{(4)}(t,\textbf{x})=&&\int
d^3\mathbf{x}'\mathcal{G}(\mathbf{x},\mathbf{x}')\biggl\{\frac{m^2\mathcal{X}}{f'(0)}T^{(2)}(t,\mathbf{x}')-g^{(2)}_
{mn,m}(t,\textbf{x}')R^{(2)}_{,n}(t,\textbf{x}')-g^{(2)}_{mn}(t,\textbf{x}')R^{(2)}_{,mn}(t,\textbf{x}')
\nonumber\\\nonumber\\
&&+R^{(2)}_{,tt}(t,\textbf{x}')-\frac{m^2}{\mu^4}\biggl[|\nabla_{\textbf{x}'}R^{(2)}(t,\textbf{x}')|^2+R^{(2)}
(t,\textbf{x}')\triangle_{\textbf{x}'}R^{(2)}(t,\textbf{x}')\biggr]
\nonumber\\\nonumber\\&&-\frac{1}{2}\nabla_{\textbf{x}'}\biggl[g^{(2)}_{tt}(t,\textbf{x}')-g^{(2)}_{mm}(t,\textbf{x}')
\biggr]\cdot
\nabla_{\textbf{x}'}R^{(2)}(t,\textbf{x}') \biggr\}
\end{eqnarray}
where $\mu^4\doteq-\frac{f'(0)}{3f'''(0)}$. Also in this case we
can have a non-vanishing curvature in absence of matter. The
solution for $g^{(4)}_{tt}$, from the first line of
(\ref{PPN-field-equation-general-theory-fR-O4}), is

\begin{eqnarray}\label{temp_temp_quarto}
g_{tt}^{(4)}(t,\textbf{x})=&&\int
d^3\mathbf{x}'\frac{1}{|\mathbf{x}-\mathbf{x}'|}\biggl\{-\frac{\mathcal{X}T^{(2)}_{tt}(t,\textbf{x}')}{2\pi
f'(0)}+\frac{1}{6\pi\mu^4}\biggl[|\nabla
R^{(2)}(t,\textbf{x}')|^2+R^{(2)}(t,\textbf{x}')\triangle
R^{(2)}(t,\textbf{x}')\biggr]
\nonumber\\\nonumber\\
&&+\frac{1}{4\pi}\biggl[g^{(2)}_{mn}(t,\textbf{x}')g^{(2)}_{tt,mn}(t,\textbf{x}')-g^{(2)}_{tt,tt}(t,\textbf{x}')
-|\nabla_{\textbf{x}'}g^{(2)}_{tt}(t,\textbf{x}')|^2-R^{(4)}(t,\textbf{x}')-g^{(2)}_{tt}(t,\textbf{x}')R^{(2)}
(t,\textbf{x}')\biggr]
\nonumber\\\nonumber\\
&& +\frac{1}{6\pi
m^2}\biggl[\frac{{R^{(2)}}^2(t,\textbf{x}')}{4}-\frac{R^{(2)}(t,\textbf{x}')\triangle
g^{(2)}_{tt}(t,\textbf{x}')}{2}+g^{(2)}_{mn,m}(t,\textbf{x}'){R^{(2)}}_{,n}
(t,\textbf{x}')+\triangle R^{(4)}(t,\textbf{x}')
\nonumber\\\nonumber\\
&&+g^{(2)}_{tt}(t,\textbf{x}')\triangle R^{(2)}(t,\textbf{x}')
+g^{(2)}_{mn}(t,\textbf{x}'){R^{(2)}}_{,mn}(t,\textbf{x}')-\frac{1}{2}\nabla
g^{(2)}_{mm}(t,\textbf{x}')\cdot\nabla
R^{(2)}(t,\textbf{x}')\biggr]\biggr\}
\end{eqnarray}

We conclude this paragraph by having shown the more general solution of field equations of $f(R)$-gravity in the
Newtonian and post-Newtonian limit assuming a coordinates transformation for the whose the gauge harmonic
condition is verified. In the next paragraph we shall apply a such scheme to obtain the explicit form of metric
tensor for a static and spherically symmetric matter source.

\section{The spacetime generated by an extended uniform mass ball-like source}\label{ball_like}

Let us consider a ball-like source with mass $M$ and radius $\xi$.
The energy-momentum tensor is (we are not interesting to the internal structure)

\begin{eqnarray}\label{emtensor}
\left\{\begin{array}{ll}T_{\mu\nu}\,=\,\rho(\mathbf{x})u_\mu
u_\nu
\\\\
T\,=\,\rho(\mathbf{x})\end{array}\right.
\end{eqnarray}
where $\rho(\mathbf{x})$ is the mass density and $u_\mu$ satisfies
the condition $g^{tt}{u_t}^2\,=\,1$, $u_i\,=\,0$. Since (\ref{metric_tensor_PPN})
the expression (\ref{emtensor})
becomes

\begin{eqnarray}\label{emtensorPPN}
\left\{\begin{array}{ll}T_{tt}(t,\mathbf{x})\,\sim\,\rho(\mathbf{x})+\rho(\mathbf{x})g^{(2)}_{tt}(t,\mathbf{x})\,
=\,T^{(0)}_{tt}(t,\mathbf{x})+T^{(2)}_
{tt}(t,\mathbf{x})
\\\\
T\,=\,\rho(\mathbf{x})\,=\,T^{(0)}(t,\mathbf{x})\end{array}\right.
\end{eqnarray}
The possible choices of Green function, for spherically symmetric systems (\emph{i.e.} $\mathcal{G(\mathbf{x},
\mathbf{x}')}\,=\,\mathcal{G}(|\mathbf{x}-\mathbf{x}'|)$), are the following

\begin{eqnarray}\label{green_function}
\mathcal{G}(\mathbf{x},\mathbf{x}')\,=\,\left\{\begin{array}{ll}-\frac{1}{4\pi}\frac{e^{-m|\mathbf{x}-\mathbf{x}'|}}
{|\mathbf{x}-\mathbf{x}'|}\,\,\,\,\,\,\,\,\,\,\,\,\,\,\,\,\,\,\,\,\,\,\,\,\,\,\,\,\,\,\,\,\,\,\,\,\,\,\,\,\,\,\,\,\,
\,\,\,\text{if}\,\,\,\,\,\,\,\,\,\,\,\,\,\,m^2\,\,>\,0
\\\\
C_1\frac{e^{-im|\mathbf{x}-\mathbf{x}'|}}{|\mathbf{x}-\mathbf{x}'|}+C_2\frac{e^{im|\mathbf{x}-\mathbf{x}'|}}
{|\mathbf{x}-\mathbf{x}'|}
\,\,\,\,\,\,\,\,\,\,\,\,\,\,\,\text{if}\,\,\,\,\,\,\,\,\,\,\,\,\,\,m^2\,<\,0\end{array}\right.
\end{eqnarray}
with $C_1+C_2\,=\,-\frac{1}{4\pi}$. It notes, for any function of
only modulus $h(|\mathbf{x}|)$, that

\begin{eqnarray}\label{int1}I\,=\,\int
d^3\mathbf{x}'\mathcal{G}(\mathbf{x},\mathbf{x}')h(\textbf{x}')\,=\,-\frac{1}{4\pi}\int
d|\mathbf{x}'||\mathbf{x}'|^2h(|\textbf{x}'|)\int_0^{2\pi}
d\phi'\int_0^\pi
d\theta'\frac{\sin\theta'e^{-m\sqrt{|\mathbf{x}|^2+|\mathbf{x}'|^2-2|\mathbf{x}||\mathbf{x}'|\cos\alpha}}}
{\sqrt{|\mathbf{x}|^2+|\mathbf{x}'|^2-2|\mathbf{x}||\mathbf{x}'|\cos\alpha}}
\end{eqnarray}
where
$\cos\alpha=\cos\theta\cos\theta'+\sin\theta\sin\theta'\cos(\phi-\phi')$
and $\alpha$ is the angle between two vectors $\mathbf{x}$,
$\mathbf{x}'$. In the spherically symmetric case we can choose
$\theta=0$ without losing generality (the symmetry of system is independent by the angle). By making the angular
integration we get

\begin{eqnarray}\label{int2}
I\,=\,-\frac{1}{2m|\textbf{x}|}\int
d|\mathbf{x}'||\mathbf{x}'|h(|\mathbf{x}'|)\biggl[e^{-m||\textbf{x}|-|\textbf{x}'||}-e^{-m(|\textbf{x}|+|\textbf{x}'|)}
\biggr]
\end{eqnarray}
An analogous relation is useful also for the Green function of Newtonian mechanics
$|\mathbf{x}-\mathbf{x}'|^{-1}$:

\begin{eqnarray}\label{int3}\int
d^3\mathbf{x}'\frac{h(|\textbf{x}'|)}{|\mathbf{x}-\mathbf{x}'|}\,=\,-\frac{2\pi}{|\mathbf{x}|}\int
d|\mathbf{x}'||\mathbf{x}'|\biggl[||\mathbf{x}|-|\mathbf{x}'||-|\mathbf{x}|-|\mathbf{x}'|\biggr]h(|\textbf{x}'|)\end{eqnarray}

\subsection{Solutions at $\mathcal{O}(2)$\,- and $\mathcal{O}(3)$\,-\,order}

Supposing that $m^2\,>\,0$ (\emph{i.e.}
$\text{sign}[f'(0)]\,=\,-\,\text{sign}[f''(0)]$) the Ricci scalar
(\ref{scalar_ricci_sol_gen}), if we hypothesize a matter density
as
$\rho(\mathbf{x})\,=\,\frac{3M}{4\pi\xi^3}\Theta(\xi-|\mathbf{x}|)$,
is

\begin{eqnarray}\label{sol_ric}
R^{(2)}(t,\textbf{x})\,=\,-\frac{3r_g}{\xi^3}\,
\biggl[1-e^{-m\xi}(1+m\,\xi)\frac{\sinh m|\mathbf{x}|}{m|\mathbf{x}|}\biggr]\Theta(\xi-|\mathbf{x}|)
-r_g\,m^2F(\xi)\frac{e^{-m|\mathbf{x}|}}{|\mathbf{x}|}\,\Theta(|\mathbf{x}|-\xi)
\end{eqnarray}
where $\Theta$ is the Heaviside function, $F(x)\,\doteq\,3\frac{mx\cosh m x-\sinh
m x}{m^3x^3}$ and $r_g\,=\,2GM$ is the Schwarzschild
radius\footnote{we have set for simplicity $f'(0)\,=\,1$ (otherwise we have to renormalize the coupling constant
$\mathcal{X}$ in the action
(\ref{HOGaction})).}. The solutions of (\ref{new_sol}),
(\ref{postnew_sol_ti}) and (\ref{post_new_ij}), given the relations (\ref{int2}) and (\ref{int3}), respectively are

\begin{eqnarray}\label{sol_pot}
g^{(2)}_{tt}(t,\mathbf{x})\,=\,-r_g\,
\biggl[\frac{3}{2\xi}+\frac{1}{m^2\xi^3}-\frac{|\textbf{x}|^2}{2\xi^3}-\frac{e^{-m\xi}(1+m\,\xi)}{m^2\xi^3}\frac{\sinh m|\mathbf{x}|}{m|\mathbf{x}|}\biggr]\,\Theta(\xi-|\mathbf{x}|)-r_g\biggl[\frac{1}{|\textbf{x}|}+\frac{F(\xi)}{3}
\frac{e^{-m|\mathbf{x}|}}{|\mathbf{x}|}\biggr]\Theta(|\mathbf{x}|-\xi)
\end{eqnarray}

\begin{eqnarray}
g^{(3)}_{ti}(t,\textbf{x})\,=\,0
\end{eqnarray}

\begin{eqnarray}\label{gsol2}
g^{(2)}_{ij}(t,\textbf{x})=-&r_g&\biggl\{\biggl[\frac{3}{2\xi}-\frac{5}{3m^2\xi^3}-\frac{|\textbf{x}|^2}{2\xi^3}+
\frac{(1+m\xi)e^{-m\xi}}{3m^2\xi^3}\biggl(2F(\textbf{x})+3\frac{\sinh m|\textbf{x}|}{m|\textbf{x}|}\biggr)\biggr]\Theta(\xi-|\mathbf{x}|)\nonumber\\\nonumber\\&&\,\,\,\,
+\biggl[\frac{1}{|\textbf{x}|}-\frac{2}
{3m^2|\textbf{x}|^3}-\frac{F(\xi)}{3}\biggl(\frac{1}{|\mathbf{x}|}-\frac{2}{m|\textbf{x}|^2}-\frac{2}
{m^2|\textbf{x}|^3}\biggr)e^{-m|\textbf{x}|}\biggr]\Theta(|\mathbf{x}|-\xi)\biggr\}
\delta_{ij}\nonumber\\\nonumber\\
-&r_g&\biggl\{\biggl[\frac{2(1+m\xi)e^{-m\xi}}{m^2\xi^3}\biggl(\frac{\sinh m|\textbf{x}|}{m|\textbf{x}|}-F(\textbf{x})\biggr)\biggr]\Theta(\xi-|\textbf{x}|)\nonumber\\\nonumber\\
&&\,\,\,\,+\biggl[
\frac{2}{m^2|\textbf{x}|^3}-\frac{2F(\xi)}{3}\biggl(
\frac{1}{|\textbf{x}|}
+\frac{3}{m|\textbf{x}|^2}
+\frac{3}{m^2|\textbf{x}|^3}\biggr)e^{-m|\textbf{x}|}
\biggr]\Theta(|\textbf{x}|-\xi)\biggr\}\frac{x_ix_j}{|\textbf{x}|^2}
\end{eqnarray}

For fixed values of the distance $|\mathbf{x}|$, the solutions
$g^{(2)}_{tt}$ and $g^{(2)}_{ij}$ depend on the value of the
radius $\xi$, then the Gauss theorem does not work also if the
Bianchi identities hold \cite{Stelle:1978}. In other words, since
the Green function does not scale as the inverse of distance but
has also an exponential behavior, the Gauss theorem is not
verified. We can affirm: \emph{the potential does not depend only
on the total mass but also on the mass - distribution in the
space}.

\begin{figure}[htbp]
  \centering
  \includegraphics[scale=1]{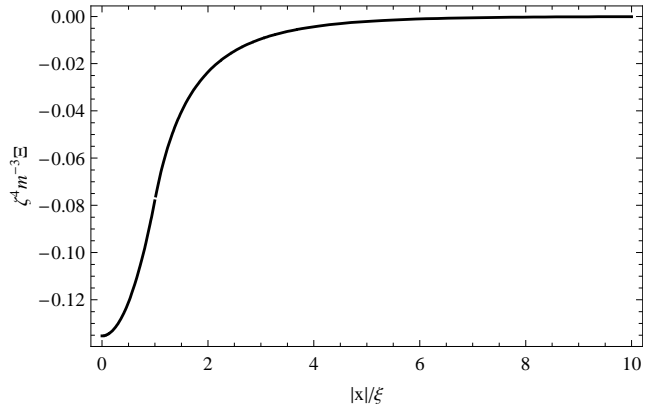}\\
  \caption{Plot of dimensionless function $\zeta^4m^{-3}\Xi$ for $\zeta\,\doteq\,m\xi\,=\,.5$ representing the
  spatial behavior of Ricci scalar at second order. In GR we would have $\Xi(\textbf{x})\,=\,\frac{3}{\xi^3}\Theta(\xi-|\mathbf{x}|)$.}
  \label{plotricciscalar}
\end{figure}

\begin{figure}[htbp]
  \centering
  \includegraphics[scale=1]{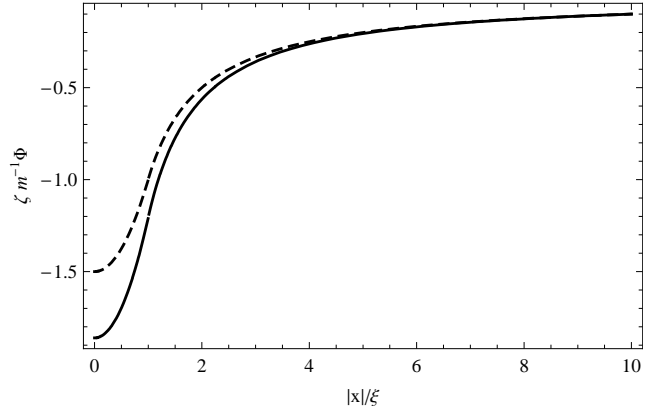}\\
  \caption{Plot of metric potential $\zeta m^{-1}\Phi$ vs distance from central mass with $\zeta\,\doteq\,m\xi\,=\,.5$.
  The dashed line is the GR behavior: $\Phi\,=\,-\biggl[\frac{3}{2\xi}-\frac{|\textbf{x}|^2}{2\xi^3}\biggr]
  \Theta(\xi-\textbf{x})-\frac{\Theta(\textbf{x}-\xi)}{|\textbf{x}|}$.}
\label{plotpontential00}
\end{figure}

\begin{figure}[htbp]
  \centering
  \includegraphics[scale=1]{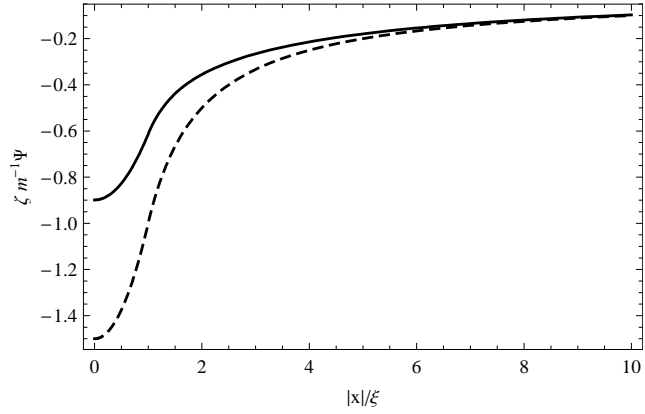}\\
  \caption{Plot of metric potential $\zeta m^{-1}\Psi$ vs distance from central mass with $\zeta\,\doteq\,m\xi\,=\,.5$.
  The dashed line is the GR behavior (similar to metric potential $\Phi$).}
\label{plotpontentialij1}
\end{figure}

\begin{figure}[htbp]
  \centering
  \includegraphics[scale=1]{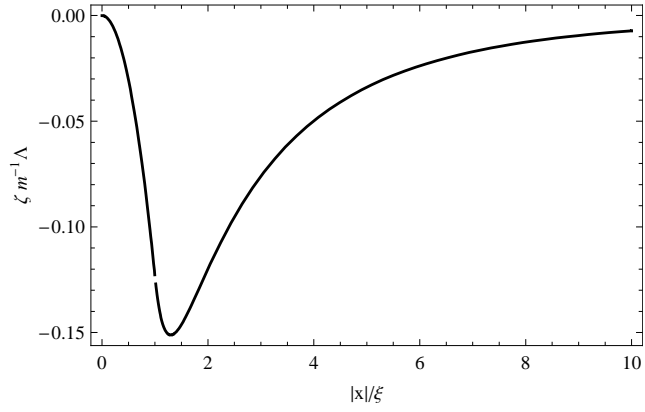}\\
  \caption{Plot of metric potential $\zeta m^{-1}\Lambda$ vs distance from central mass with $\zeta\,\doteq\,m\xi\,=\,.5$.
  In GR a such behavior is missing.}
\label{plotpontentialij2}
\end{figure}

By introducing three metric potentials $\Phi(\textbf{x})$,
$\Psi(\textbf{x})$ and $\Lambda(\textbf{x})$ (the dimension is the
inverse of length) we can rewrite (\ref{sol_pot}) and
(\ref{gsol2}) as follows

\begin{eqnarray}\label{PAS}
\left\{\begin{array}{ll}
g^{(2)}_{tt}(t,\textbf{x})\,=\,r_g\Phi(\textbf{x})
\\\\
g^{(2)}_{ij}(t,\textbf{x})\,=\,r_g\Psi(\textbf{x})\delta_{ij}+r_g\Lambda(\textbf{x})\frac{x_ix_j}{|\mathbf{x}|^2}
\\\\
\end{array}\right.
\end{eqnarray}
and with a fourth function, $\Xi(\textbf{x})$, (the dimension is the cubic inverse of length) the Ricci scalar
(\ref{sol_ric}) is

\begin{eqnarray}\label{PAS_1}
R^{(2)}(t,\textbf{x})\,=\,r_g\Xi(\textbf{x})
\end{eqnarray}
The spatial behavior (\ref{sol_ric}) is shown in FIG.
\ref{plotricciscalar}. The metric potentials are shown in FIGs.
\ref{plotpontential00}, \ref{plotpontentialij1} and
\ref{plotpontentialij2}.
It is interesting to note as the function $\Phi$ assumes
smaller value of its equivalent in GR, then in terms of
gravitational attraction we have a potential well more deep. A
such scheme can be interpretable or assuming a variation of the
gravitational constant $G$ or requiring that there is a central
greater mass. These two affirmations are compatible on the one
hand with the tensor-scalar theories (in the which we have a
scaling of gravitational constant) and on the other hand with the
theory of GR plus the hypothesis of the existence of the dark
matter. In particular, if the mass distribution takes a bigger
volume, the potential increases and vice versa.


In the limit of point-like source, \emph{i.e.}
$\lim_{\xi\rightarrow 0}\frac{3M}{3\pi
\xi^3}\Theta(\xi-|\mathbf{x}|)\,=\,M\delta(\mathbf{x})$ and
$\lim_{x\rightarrow 0}F(x)\,=\,1$, we get

\begin{eqnarray}\label{sol_new_p}
\left\{\begin{array}{ll} R^{(2)}(t,\textbf{x})=
-r_gm^2\frac{e^{-m|\mathbf{x}|}}{|\mathbf{x}|}
\\\\
g^{(2)}_{tt}(t,\mathbf{x})=-r_g\biggl(\frac{1}{|\textbf{x}|}+\frac{1}{3}\frac{e^{-m|\mathbf{x}|}}{|\mathbf{x}|}\biggr)
\\\\
g^{(2)}_{ij}(t,\textbf{x})=-r_g\biggl\{\frac{1}{|\mathbf{x}|}-\frac{2}{3m^2|\textbf{x}|^3}
-\frac{1}{3}\biggl(\frac{1}{|\mathbf{x}|}-\frac{2}{m|\textbf{x}|^2}-\frac{2}
{m^2|\textbf{x}|^3}\biggr)e^{-m
|\mathbf{x}|}\biggr\}\delta_{ij}
\\\\\,\,\,\,\,\,\,\,\,\,\,\,\,\,\,\,\,\,\,\,\,\,\,\,\,\,\,\,\,\,\,-r_g\biggl[\frac{2}{m^2|\textbf{x}|^3}-\frac{2}{3}
\biggl(\frac{1}{|\textbf{x}|}
+\frac{3}{m|\textbf{x}|^2}
+\frac{3}{m^2|\textbf{x}|^3}\biggr)e^{-m|\textbf{x}|}\biggr]\frac{x_ix_j}{|\textbf{x}|^2}
\end{array}\right.
\end{eqnarray}

An important point has to be considered. The PPN-parameters
$\gamma$ and $\beta$, in the GR context, are intended to
parameterize the deviations from the Newtonian behavior of the
gravitational potentials. They are defined according to the
standard Eddington metric (in the vacuum)

\begin{eqnarray}\label{PPN_parameters}
\left\{\begin{array}{ll}
g_{tt}=1-\frac{r_g}{|\mathbf{x}|}+\frac{\beta}{2}\frac{{r_g}^2}{|\mathbf{x}|^2}
\\\\
g_{ij}=-\delta_{ij}-\gamma\frac{r_g}{|\mathbf{x}|}\delta_{ij}
\end{array}\right.
\end{eqnarray}
In particular, the PPN parameter $\gamma$ is related with the
second order correction to the gravitational potential while
$\beta$ is linked with the fourth order perturbation in $v$. Since
the Gauss theorem is not verified for $f(R)$-gravity, while the
relations (\ref{PPN_parameters}) satisfy it, we must consider the
relations (\ref{sol_new_p}) and not (\ref{sol_pot}),
(\ref{gsol2}). After making the limit $f\rightarrow R$ we can
compare the results with (\ref{PPN_parameters}). Actually, if we
consider this limit for (\ref{sol_new_p}), we have

\begin{eqnarray}\label{sol_GR_vacuum}
\left\{\begin{array}{ll} R^{(2)}(t,\textbf{x})\,=\,0
\\\\
g^{(2)}_{tt}(t,\mathbf{x})\,=\,-\frac{r_g}{|\textbf{x}|}
\\\\
g^{(2)}_{ij}(t,\textbf{x})\,=\,-\frac{r_g}{|\mathbf{x}|}\delta_{ij}
\end{array}\right.
\end{eqnarray}
which suggest that the $f(R)$-gravity is compatible with respect
to GR. At last it is interesting to note that also in the case of
extended spherically symmetric distribution of matter when we
perform the limit $f\rightarrow R$ the solutions (\ref{sol_pot})
and (\ref{gsol2}) directly converge (in the vacuum) in solutions
(\ref{sol_GR_vacuum}), demonstrating the validity of Gauss theorem
in GR.

Other important consideration is about the asymptotic behavior of
$f$-theory with respect to GR. In fact increasing the distance
from the central mass the gravitational field brings near to one
of GR. A such rejoining is a normal consequence of the hypothesis
(spherically symmetric system and request of Minkowskian limit).
At last in FIG. \ref{plotricciscalar} we report the spatial
behavior of Ricci scalar (\ref{sol_ric}) approximating
asymptotically the given value in GR. In fact hypothesizing a
$f(R)$- theory the Ricci scalar acquires dynamics, and in the
Newtonian limit, we find a characteristic scale length ($m^{-1}$)
on the which distance the scalar massive field evolves. Only for
distances more great than $m^{-1}$ we recover the outcome of GR:
$R\,=\,0$.

\begin{figure}[htbp] \centering
  \includegraphics[scale=1]{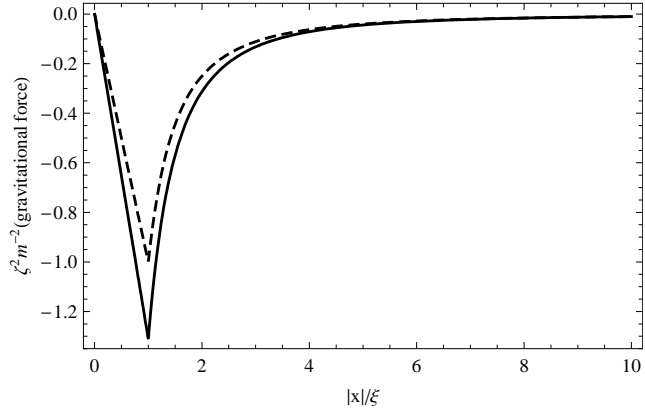}\\
  \caption{Comparison between gravitational forces induced by GR and $f(R)$-theory with $\zeta\,\doteq\,m\xi\,=\,.5$.
  The dashed line is the GR behavior.}
  \label{plotforce}
\end{figure}

To conclude this section we show in FIG. \ref{plotforce} the
comparison between gravitational forces induced in GR and in
$f(R)$-theory in the Newtonian limit. Obviously also about the
force we obtained an intensity stronger than in GR.

\subsection{The oscillating Newtonian Limit of $f(R)$-gravity}

If we consider $m^2\,<\,0$ (i.e.
$\text{sign}[f'(0)]\,=\,\text{sign}[f''(0)]$) from
(\ref{green_function}) we can choose the "oscillating" Green
function

\begin{eqnarray}\label{green_function_2}
\mathcal{G}(\mathbf{x},\mathbf{x}')\,=\,-\frac{1}{4\pi}\frac{\cos
m|\mathbf{x}-\mathbf{x}'|+ \sin
m|\mathbf{x}-\mathbf{x}'|}{|\mathbf{x}-\mathbf{x}'|}
\end{eqnarray}
The Ricci scalar (\ref{scalar_ricci_sol_gen}) and the
$tt$-component of $g_{\mu\nu}$ at $\mathcal{O}(2)$ order
(\ref{new_sol}) become

\begin{eqnarray}\label{sol_ric_oscil}
R^{(2)}(t,\textbf{x})=\,-\frac{6r_g}{\xi^3}\,\biggl[1-H(\xi)\frac{\sin
m|\mathbf{x}|}{m|\mathbf{x}|}\biggr]\Theta(\xi-|\mathbf{x}|)
-2\,r_g\,m^2G(\xi)\frac{\cos m|\textbf{x}|+\sin
m|\textbf{x}|}{|\mathbf{x}|}\,\Theta(|\mathbf{x}|-\xi)
\end{eqnarray}

\begin{eqnarray}\label{sol_pot_oscil}
g^{(2)}_{tt}(t,\mathbf{x})\,=\,-r_g\,
\biggl[\frac{3}{2\xi}-\frac{2}{m^2\xi^3}-\frac{|\textbf{x}|^2}{2\xi^3}+\frac{2H(\xi)}{m^2\xi^3}\frac{\sin
m|\mathbf{x}|}{m|\mathbf{x}|}\biggr]\,\Theta(\xi-|\mathbf{x}|)-r_g\biggl[\frac{1}{|\textbf{x}|}-\frac{2G(\xi)}{3}
\frac{\cos m|\mathbf{x}|+\sin
m|\mathbf{x}|}{|\mathbf{x}|}\biggr]\Theta(|\mathbf{x}|-\xi)
\end{eqnarray}
where $G(\xi)\,\doteq\,3\frac{m\xi\cos m\xi-\sin m\xi}{m^3\xi^3}$
and $H(\xi)\,\doteq\,(1-m\xi)\cos m\xi+(1+m\xi)\sin m\xi$, with
the properties $\lim_{\xi\rightarrow 0}G(\xi)\,=\,-1$ and
$\lim_{\xi\rightarrow 0}H(\xi)\,=\,1$. Since we have an
oscillating Green function (it is not asymptotically zero) the
"gravitational potentials" (\ref{sol_pot_oscil}) at infinity are
zero unless a constant value
($\lim_{a\rightarrow\infty}2r_g\,m\,G(\xi)(\sin ma-\cos ma)$).

\begin{figure}[htbp]
  \centering
  \includegraphics[scale=1]{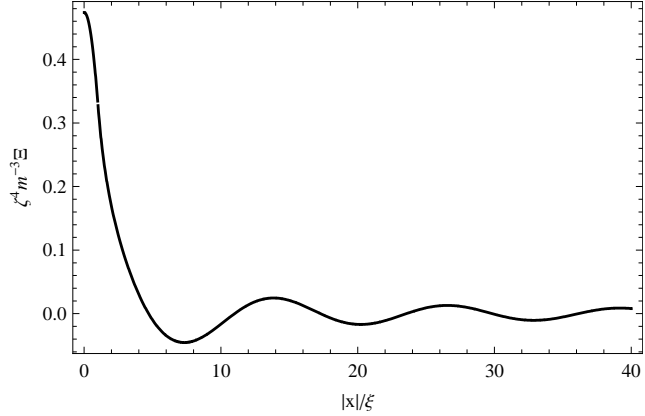}\\
  \caption{Plot of dimensionless function $\zeta^4m^{-3}\Xi$ with $\zeta\,\doteq\,m\xi\,=\,.5$ representing the
  spatial behavior of Ricci scalar at second order in the oscillating case.}
  \label{plotricciscalar_oscil}
\end{figure}

\begin{figure}[htbp]
  \centering
  \includegraphics[scale=1]{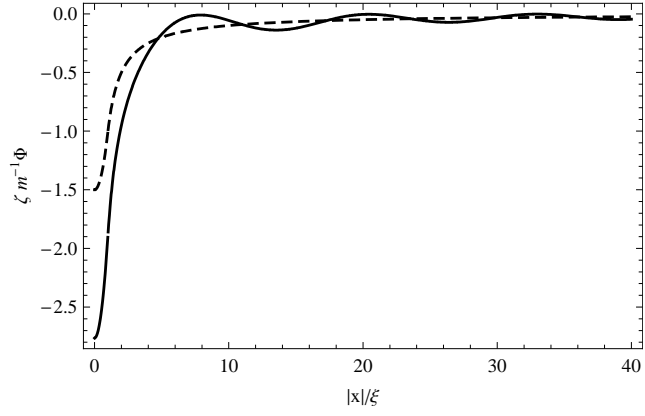}\\
  \caption{Plot of metric potential $\zeta m^{-1}\Phi$ vs distance from central mass with the choice
  $\zeta\,\doteq\,m\xi\,=\,.5$ in the oscillating case. The dashed line is the GR behavior.}
\label{plotpontential00_oscil}
\end{figure}

The spatial behavior of Ricci scalar (\ref{sol_ric_oscil}) and
metric component (\ref{sol_pot_oscil}) are shown in FIGs.
\ref{plotricciscalar_oscil} and \ref{plotpontential00_oscil}.
The considerations of preceding subsection hold also for the
solutions (\ref{sol_ric_oscil}) - (\ref{sol_pot_oscil}). The only
difference is that now we have oscillating behaviors instead of
exponential behaviors. The correction term to the Newtonian
potential in the external solution can be interpreted as the
Fourier transform of the matter density $\rho(\mathbf{x})$. In
fact, we have:

\begin{eqnarray}
\lim_{\xi\rightarrow 0}\int d^3\mathbf{x}'\rho(\mathbf{x}')e^{-i\mathbf{k}\cdot\mathbf{x}'}\,=\,-\lim_{\xi\rightarrow 0}M\,G(|\mathbf{k}|\xi)\,=\,M
\end{eqnarray}

\begin{figure}[htbp] \centering
  \includegraphics[scale=1]{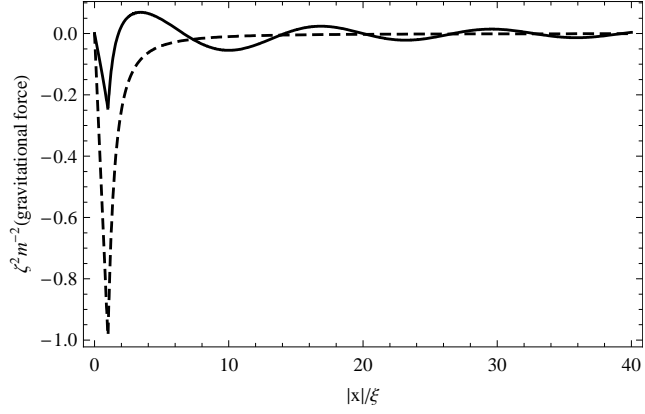}\\
  \caption{Comparison between gravitational forces induced by GR and $f(R)$-theory with $\zeta\,\doteq\,m\xi\,=\,.5$ in
  the oscillating case. The dashed line is the GR behavior.}
  \label{plotforce_oscil}
\end{figure}

Also in this case we conclude the section showing in FIG.
\ref{plotforce_oscil} the comparison between gravitational forces
induced in GR and in $f(R)$-theory in the Newtonian limit.
Obviously also in this last case we obtained a force stronger than
in GR.

\subsection{Solutions at $\mathcal{O}(4)$\,-\,order}

The metric potentials and the function $\Xi(\mathbf{x})$, respectively defined in (\ref{PAS}) and (\ref{PAS_1}), satisfy the
following properties with respect to derivative of coordinate $l$-th\footnote{we remember that $|\textbf{x}|_{,l}\,=\,|\textbf{x}|^{-1}\,x_l$} in the matter

\begin{eqnarray}\label{relations_matter}
\left\{\begin{array}{ll}
\Xi_{,l}(\textbf{x})\,=\,\frac{m^2(1+m\xi)}{\xi^3}\,e^{-m\xi}\,F(\mathbf{x})\,x_l\,=\,
\Xi_0(\textbf{x})\,x_l
\\\\
\Phi_{,l}(\textbf{x})\,=\,\biggl[\frac{1}{\xi^3}+\frac{(1+m\xi)}{3\xi^3}\,e^{-m\xi}\,F(\mathbf{x})\biggr]\,x_l\,=\,\Phi_0(\textbf{x})\,x_l
\\\\
\Psi_{,l}(\textbf{x})\,=\,\biggl[\frac{1}{\xi^3}-\frac{(1+m\xi)}{\xi^3}\,e^{-m\xi}\,\frac{(m^2|\textbf{x}|^2+6)\sinh
m|\mathbf{x}|+m|\textbf{x}|(m^2|\textbf{x}|^2-6)\cosh
m|\textbf{x}|}{m^5|\textbf{x}|^5}\biggr]\,x_l\,=\,\Psi_0
(\textbf{x})\,x_l
\\\\
\Lambda_{,l}(\textbf{x})\,=\,\frac{2(1+m\xi)}{\xi^3}\,e^{-m\xi}\,\frac{(4m^2|\textbf{x}|^2+9)\sinh
m|\mathbf{x}|-m|\textbf{x}|(m^2|\textbf{x}|^2+9)\cosh
m|\textbf{x}|}{m^5|\textbf{x}|^5}\,x_l\,=\,\Lambda_0(\textbf{x})
\,x_l\\\\
\Xi_{,ln}(\textbf{x})\,=\,\Xi_0(\textbf{x})\delta_{ln}+\frac{3(1+m\xi)}{\xi^3}\,e^{-m\xi}\,\frac{(m^2|\mathbf{x}|^2+3)\sinh
m |\mathbf{x}|-3m|\mathbf{x}|\cosh
m|\mathbf{x}|}{m|\mathbf{x}|^5}\,x_lx_n\,=\,\Xi_0(\textbf{x})\delta_{ln}+\Xi_1(\textbf{x})x_lx_n
\\\\
\Phi_{,ln}(\textbf{x})\,=\,\Phi_0(\textbf{x})\delta_{ln}+\frac{(1+m\xi)}{\xi^3}\,e^{-m\xi}\,\frac{(m^2|\mathbf{x}|^2+3)\sinh
m |\mathbf{x}|-3m|\mathbf{x}|\cosh
m|\mathbf{x}|}{m^3|\mathbf{x}|^5}
\,x_lx_n\,=\,\Phi_0(\textbf{x})\delta_{ln}+\Phi_1(\textbf{x})x_lx_n
\end{array}\right.
\end{eqnarray}
and in the vacuum

\begin{eqnarray}\label{relations}
\left\{\begin{array}{ll}
\Xi_{,l}(\textbf{x})\,=\,\frac{m^2(m|\textbf{x}|+1)}{|\textbf{x}|^3}\,F(\xi)\,e^{-m|\textbf{x}|}\,x_l\,=\,
\Xi_0(\textbf{x})\,x_l
\\\\
\Phi_{,l}(\textbf{x})\,=\,\biggl[\frac{1}{|\textbf{x}|^3}+\frac{m|
\textbf{x}|+1}{|\textbf{x}|^3}\frac{F(\xi)e^{-m|\textbf{x}|}}{3}\biggr]\,x_l\,=\,\Phi_0(\textbf{x})\,x_l
\\\\
\Psi_{,l}(\textbf{x})\,=\,\biggl[\frac{m^2|\textbf{x}|^2-2}{m^2|\textbf{x}|^5}-\frac{m^3
|\textbf{x}|^3-m^2|\textbf{x}|^2-6m|\mathbf{x}|-6}{m^2|\textbf{x}|^5}\frac{F(\xi)e^{-m|\textbf{x}|}}{3}\biggr]\,x_l\,=\,\Psi_0
(\textbf{x})\,x_l
\\\\
\Lambda_{,l}(\textbf{x})\,=\,\biggl[\frac{6}{m^2|\textbf{x}|^5}-\frac{m^3|\textbf{x}|
^3+4m^2|\textbf{x}|^2+9m|\textbf{x}|+9}{m^2|\textbf{x}|^5}\frac{2F(\xi)e^{-m|\textbf{x}|}}{3}\biggr]\,x_l\,=\,\Lambda_0(\textbf{x})
\,x_l\\\\
\Xi_{,ln}(\textbf{x})\,=\,\Xi_0(\textbf{x})\delta_{ln}-\frac{m^2(m^2|\textbf{x}|^2+3m|\textbf{x}|+3)}{
|\textbf{x}|^5}F(\xi)e^{-m|\textbf{x}|}x_lx_n\,=\,\Xi_0(\textbf{x})\delta_{ln}+\Xi_1(\textbf{x})x_lx_n
\\\\
\Phi_{,ln}(\textbf{x})\,=\,\Phi_0(\textbf{x})\delta_{ln}-\biggl[\frac{3}{|
\textbf{x}|^5}+\frac{m^2|\textbf{x}|^2+3m|\textbf{x}|+3}{|\textbf{x}|^5}\frac{F(\xi)e^{-m|\textbf{x}|}}{3}\biggr]x_lx_n\,=\,\Phi_0(\textbf{x})\delta_{ln}+\Phi_1(\textbf{x})x_lx_n
\end{array}\right.
\end{eqnarray}
Obviously when we consider the physics in the matter or in the
vacuum we choose the "right" quantities $\Xi_0(\textbf{x})$,
$\Xi_1(\textbf{x})$, $\Phi_0(\textbf{x})$, $\Phi_1(\textbf{x})$,
$\Psi_0(\textbf{x})$, $\Lambda_0(\textbf{x})$.

The expression of Ricci scalar at fourth order,
(\ref{Ricci_quarto}), is

\begin{eqnarray}\label{Ricci_quarto_solution}
R^{(4)}(t,\textbf{x})=&&\frac{{r_g}^2}{2m|\textbf{x}|}\int_0^\infty
d|\mathbf{x}'||\mathbf{x}'|\biggl\{e^{-m||\textbf{x}|-|\textbf{x}'||}-e^{-m(|\textbf{x}|+|\textbf{x}'|)}\biggr\}\biggl
\{\frac{m^4}{\mu^4}\biggl[\Xi(\textbf{x}')^2+\frac{|\textbf{x}'|^2}{m^2}\Xi_0(\textbf{x}')^2\biggr]
\nonumber\\\nonumber\\
&&+\biggl[3\Lambda(\mathbf{x}')+\frac{\Phi_0(\mathbf{x}')-\Psi_0(\mathbf{x}')+\Lambda_0(\textbf{x}')}{2}|\textbf{x}'|^2
\biggr]\Xi_0(\textbf{x}')
+m^2\Psi(\textbf{x}')\Xi(\textbf{x}')+\Xi_1(\textbf{x}')\Lambda(\textbf{x}')|\textbf{x}'|^2\biggr\}
\end{eqnarray}
from the which we note two contributes. The first one still
depends on the quadratic term ($\propto R^2$) in the action
(\ref{HOGaction}), while the second one is related to cubic term
($\propto R^3$). By introducing two functions $\Xi_I(\textbf{x})$
and $\Xi_{II}(\textbf{x})$, the (\ref{Ricci_quarto_solution}) is
rewritable as follows

\begin{eqnarray}\label{PAS_Ricci_quarto_solution}
R^{(4)}(t,\textbf{x})={r_g}^2\biggl[\Xi_I(\textbf{x})+\frac{m^4}{\mu^4}\Xi_{II}(\textbf{x})\biggr]
\end{eqnarray}

An analogous situation is found for the $tt$-component of metric
tensor at fourth order. In fact the (\ref{temp_temp_quarto})
becomes

\begin{eqnarray}\label{g00_quarto_solution}
g_{tt}^{(4)}(t,\textbf{x})=&&\frac{r_g\mathcal{X}}{|\mathbf{x}|}\int_0^\xi
d|\mathbf{x}'||\textbf{x}'|\biggl\{||\mathbf{x}|-|\mathbf{x}'||-|\mathbf{x}|-|\mathbf{x}'|\biggr\}
\rho(\mathbf{x}')\Phi(\mathbf{x}')\nonumber\\\nonumber\\&&
+\frac{{r_g}^2}{|\mathbf{x}|}\int_0^\infty
d|\mathbf{x}'||\textbf{x}'|\biggl\{||\mathbf{x}|-|\mathbf{x}'||-|\mathbf{x}|-|\mathbf{x}'|\biggr\}\biggl\{
\frac{1}{2}\biggl[\Xi_I(\mathbf{x}')+\Phi(\mathbf{x}')\Xi(\mathbf{x}')+\Phi_0(\mathbf{x}')^2|\mathbf{x}'|^2
\biggr]\nonumber\\\nonumber\\&&
\,\,\,\,\,\,\,\,\,\,\,\,\,\,\,\,\,\,\,\,\,\,\,\,\,\,\,\,\,\,\,\,\,\,\,\,\,\,\,\,\,\,\,\,\,\,\,\,\,\,\,\,\,\,\,\,
\,\,\,\,\,\,\,\,\,\,\,\,\,\,\,
-\frac{1}{2}\biggl[\biggl(3\Psi(\mathbf{x}')+\Lambda(\mathbf{x}')\biggr)\Phi_0(\mathbf{x}')+
\biggl(\Psi(\mathbf{x}')+\Lambda(\mathbf{x}')\biggr)\Phi_1(\mathbf{x}')|\mathbf{x}'|^2\biggr]\biggr\}
\nonumber\\\nonumber\\&&
-\frac{{r_g}^2}{3m^2|\mathbf{x}|}\int_0^\infty
d|\mathbf{x}'||\textbf{x}'|\biggl\{||\mathbf{x}|-|\mathbf{x}'||-|\mathbf{x}|-|\mathbf{x}'|\biggr\}
\biggl\{\frac{\Xi(\textbf{x}')^2}{4}+\triangle_{\mathbf{x}'}\Xi_I(\mathbf{x}')+\Xi_1(\mathbf{x}')\Lambda
(\mathbf{x}')|\mathbf{x}'|^2 \nonumber\\\nonumber\\&&
\,\,\,\,\,\,\,\,\,\,\,\,\,\,\,\,\,\,\,\,\,\,\,\,\,\,\,\,\,\,\,\,\,\,\,\,\,\,\,\,\,\,\,\,\,\,\,\,\,\,\,\,\,\,\,\,
\,\,\,\,\,\,\,\,\,\,\,\,\,\,\,\,\,\,\,\,\,\,\,\,\,\,\,\,\,\,\,\,\,\,\,\,\,\,\,\,\,\,\,\,\,\,\,\,\,\,\,\,\,\,\,\,
+\frac{\Xi_0(\mathbf{x}')}{2}\biggr[6\Lambda(\mathbf{x}')+\biggl(5\Psi_0(\mathbf{x}')
+3\Lambda_0(\mathbf{x}')\biggr)|\mathbf{x}'|^2\biggr]\nonumber\\\nonumber\\&&
\,\,\,\,\,\,\,\,\,\,\,\,\,\,\,\,\,\,\,\,\,\,\,\,\,\,\,\,\,\,\,\,\,\,\,\,\,\,\,\,\,\,\,\,\,\,\,\,\,\,\,\,\,\,\,\,
\,\,\,\,\,\,\,\,\,\,\,\,\,\,\,\,\,\,\,\,\,\,\,\,\,\,\,\,\,\,\,\,\,\,\,\,\,\,\,\,\,\,
+\frac{\Xi(\mathbf{x}')}{2}\biggr[2m^2\biggl(\Phi(\mathbf{x}')+\Psi(\mathbf{x}')\biggr)-3\Phi_0(\mathbf{x}')
-\Phi_1|\mathbf{x}'|^2\biggr]\biggr\} \nonumber\\\nonumber\\&&
+\frac{m^4}{\mu^4}\frac{{r_g}^2}{|\mathbf{x}|}\int_0^\infty
d|\mathbf{x}'||\textbf{x}'|\biggl\{||\mathbf{x}|-|\mathbf{x}'||-|\mathbf{x}|-|\mathbf{x}'|\biggr\}
\biggl\{\frac{\Xi_{II}(\textbf{x}')}{2}-\frac{m^2\Xi(\textbf{x}')^2+|\textbf{x}'|^2\Xi_0(\textbf{x}')^2}{3m^4}
\nonumber\\\nonumber\\&&
\,\,\,\,\,\,\,\,\,\,\,\,\,\,\,\,\,\,\,\,\,\,\,\,\,\,\,\,\,\,\,\,\,\,\,\,\,\,\,\,\,\,\,\,\,\,\,\,\,\,\,\,\,\,\,\,
\,\,\,\,\,\,\,\,\,\,\,\,\,\,\,\,\,\,\,\,\,\,\,\,\,\,\,\,\,\,\,\,\,\,\,\,\,\,\,\,\,\,\,\,\,\,\,\,\,\,\,\,\,\,\,\,
\,\,\,\,\,\,\,\,\,\,\,\,\,\,\,\,\,\,\,\,\,\,\,\,\,\,\,\,\,\,\,\,\,\,\,\,\,\,\,\,\,\,\,\,\,\,\,\,\,\,\,\,\,\,\,\,
\,\,\,\,\,\,\,\,\,\,\,\,\,\,\,\,\,\,\,\,\,\,\,\,\,\,\,
\,\,\,-\frac{\triangle_{\mathbf{x}'}\Xi_{II}(\mathbf{x}')}{3m^2}\biggr\}
\end{eqnarray}
and by introducing other new functions $\Phi_I(\textbf{x})$, $\Phi_{II}(\textbf{x})$ we have

\begin{eqnarray}\label{PAS_g00_quarto_solution}
g^{(4)}_{tt}(t,\textbf{x})={r_g}^2\biggl[\Phi_I(\textbf{x})+\frac{m^4}{\mu^4}\Phi_{II}(\textbf{x})\biggr]
\end{eqnarray}

It is useful to note that we have generally four contributes to
$g_{tt}^{(4)}$ in (\ref{g00_quarto_solution}). The first one is
induced by the non-linearity of the metric tensor even in our
static spherically symmetric case. The product
$\rho(\mathbf{x})\Phi(\mathbf{x})$ is not zero only in the matter
but contributes to determination of $tt$-component in any point of
space. The second one holds account of the induced contribution,
by solution of previous order, to the determination of the
$tt$-component of the Ricci tensor at fourth order. These first
two terms are present also in GR. While the second two ones are
derived from the modification of theory. In fact the third
contribution depends on the addition of the quadratic term
($\propto R^2$) in the action and finally the fourth one from the
addition of the cubic term ($\propto R^3$).

\begin{figure}[htbp] \centering
  \includegraphics[scale=1]{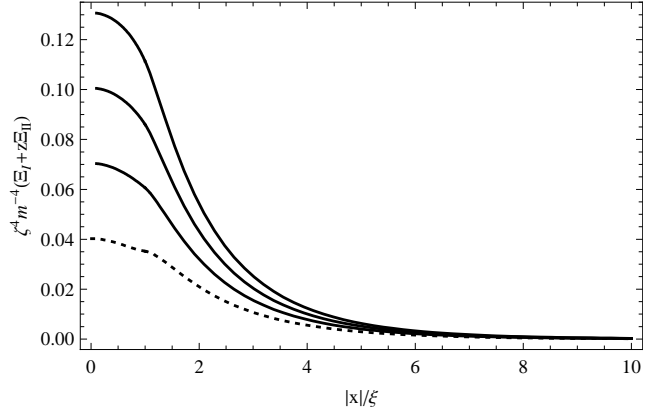}\\
  \caption{Plot of dimensionless function $\zeta^4m^{-4}(\Xi_I+z\,\Xi_{II})$, representing the Ricci scalar at fourth
  order, where $z\,=\,m^4/\mu^4$ and $\zeta\,\doteq\,m\xi\,=\,.5$. The spatial behavior is shown for $.3\,\leqslant\,z\,\leqslant\,.9$
  (solid lines) while the dotted line corresponds to $R-\frac{1}{6m^2}R^2$-theory.}
  \label{plotricciquartoordine}
\end{figure}

The choice of free parameter $\mu$ (which is linked to third
derivative of $f(R)$) is a crucial point in both the expressions
(\ref{PAS_Ricci_quarto_solution}) and
(\ref{PAS_g00_quarto_solution}) to obtain the right behavior. From
mathematical interpretation of Newtonian limit one has
$|f'''(0)|\,<\,|f''(0)|$ and if $\mu^4\,>\,0$ (\emph{i.e.}
$\text{sign}[f'(0)]\,=\,-\,\text{sign}[f'''(0)]$, otherwise
$\mu^4$ is not a length) we have
$m^4/\mu^4\,=\,|f'''(0)|/3{f''(0)}^2$, so we find the constraint
$0\,<\,m^4/\mu^4\,<\,1$. In FIG. (\ref{plotricciquartoordine}) we
report the spatial behavior of (\ref{PAS_Ricci_quarto_solution})
in the matter and in the vacuum,
($.3\,\leqslant\,m^4/\mu^4\,\leqslant\,.9$), showing that far from
source we obtain a spacetime with a vanishing scalar curvature. At
Newtonian level the Ricci scalar $R^{(2)}$ is negative defined
(\ref{sol_ric}) while at Post-Newtonian limit is positive
defined.

\begin{figure}[htbp] \centering
  \includegraphics[scale=1]{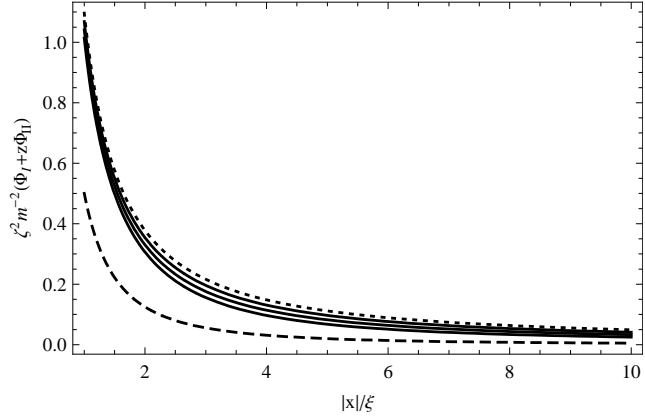}\\
  \caption{Plot of dimensionless function $\zeta^2m^{-2}(\Phi_I+z\,\Phi_{II})$ (solid lines) where $z\,=\,m^4/\mu^4$ and
  of function $1/2|\mathbf{x|^2}$ (dashed line). For $z\,=\,0$ (dotted line) we have the behavior of $R-\frac{1}{6m^2}R^2$-theory.
  The solid lines are obtained for $.3\,\leqslant\,z\,\leqslant\,.9$ and for $\zeta\,=\,m\,\xi\,=\,.5$.}
  \label{plotg00quartoordine}
\end{figure}

In FIG. (\ref{plotg00quartoordine}) we report the time-time
component of metric tensor, $g^{(4)}_{tt}$, on the same interval
of values of $m^4/\mu^4$, although the behavior is quite
insensitive to changes induced by the contributions of the cubic
term in the Lagrangian. Besides we can observe an important
analogy with respect the results of GR. In both cases we have a
potential barrier, but for $f(R)$-gravity it is higher (as in the
Newtonian limit we found a deeper potential well).


\subsection{Solutions from isotropic coordinates to standard coordinates}

The found solutions of metric are expressed in isotropic
coordinates and often for spherically symmetric problems is
conveniently rewritten in standard coordinates (the usual form in
which we write the Schwarzschild solution). Here the relativistic
invariant of metric (\ref{metric_tensor_PPN}) is

\begin{eqnarray}\label{me0}
{ds}^2\,=\,\biggl[1+r_g\Phi(\mathbf{x})+{r_g}^2\biggl(\Phi_I(\mathbf{x})+\frac{m^4}{\mu^4}\Phi_{II}(\mathbf{x})
\biggr)\biggr]dt^2-\biggl[1-r_g\Psi(\mathbf{x})\biggr]|d\mathbf{x}|^2+r_g\Lambda(\mathbf{x})\frac{(\mathbf{x}\cdot
d\mathbf{x})^2}{|\mathbf{x}|^2}
\end{eqnarray}
From spherically symmetric form of (\ref{me0}) it is convenient to
replace the position $\textbf{x}$ with spherical polar coordinates $r, \theta,
\phi$ defined as usual by

\begin{eqnarray}x^1\,=\,r\sin\theta\cos\phi\,,\,\,\,\,x^2\,=\,r\sin\theta\sin\phi\,,\,\,\,\,x^3\,=\,r\cos\phi\,.
\end{eqnarray}
The proper time interval (\ref{me0}) then becomes

\begin{eqnarray}\label{me1}
{ds}^2\,=\,\biggl[1+r_g\Phi(r)+{r_g}^2\biggl(\Phi_I(r)+\frac{m^4}{\mu^4}\Phi_{II}(r)\biggr)
\biggr]dt^2-
\biggl[1-r_g\biggl(\Psi(r)+\Lambda(r)\biggr)\biggr]dr^2-\biggl[1-r_g\Psi(r)\biggr]r^2d\Omega
\end{eqnarray}
where $d\Omega=d\theta^2+\sin^2\theta d\phi^2$ is the solid angle. To get the metric in the standard form it needs
to impose a radial coordinate transformation

\begin{eqnarray}\label{condit_transf}
\biggl[1-r_g\Psi(r)\biggr]r^2\,=\,\tilde{r}^2
\end{eqnarray}
and we have a new set of coordinates $\tilde{r},\theta,\phi$. The metric (\ref{me1}) becomes

\begin{eqnarray}\label{metricsolvedstandar}
{ds}^2\,=\,\biggl[1+r_g\tilde{\Phi}(\tilde{r})+{r_g}^2\biggl(\tilde{\Phi}_I(\tilde{r})+\frac{m^4}{\mu^4}
\tilde{\Phi}_{II}(\tilde{r})\biggr)
\biggr]dt^2-
\biggl[1-r_g\biggl(\tilde{\Psi}(\tilde{r})+\tilde{\Lambda}(\tilde{r})\biggr)\biggr]\biggr(\frac{dr}{d\tilde{r}}
\biggr)^2d\tilde{r}^2-\tilde{r}^2d\Omega
\end{eqnarray}
the expression previously desired.

The explicit expression of (\ref{metricsolvedstandar}) is not
displayed because the equation (\ref{condit_transf}) can not be
solved algebraically. However, this technical problem is overcome
with the help of numerical methods when we are interesting to test
experimentally the theory. In a further work that is taking place,
we are considering the main experiments tested for GR and we want
retest them with respect to the results obtained.

\section{Conclusions}\label{conclusions}

In this paper, we have generally reformulated the Newtonian limit
of $f(R)$-gravity by applying the Green function method. Moreover
the post-Newtonian limit has been studied in the harmonic gauge
condition and the relative spatial behaviors (Yukawa-like and
oscillating) of gravitational potential in the matter and in the
vacuum have been shown. The Taylor expansion of a generic $f$ has
been considered obtaining general solutions in term of the
derivatives up to third degree when an uniform mass ball-like
source is considered. All metric potentials, however, depend
strictly on the coupling parameters appearing indirectly in the
Lagrangian of the theory (the derivatives of $f$ in $R\,=\,0$ are
arbitrary constants).

A detailed discussion has been developed for systems presenting
spherical symmetry. In this case, the role of corrections to the
Newtonian potential is clearly evident. This means that one of the
effects introducing a generic function of scalar curvature is to
select a characteristic scale length which could have physical
interests.

Furthermore, it has been shown that the Birkhoff theorem is not a
general result for $f(R)$-gravity. This is a fundamental
difference between GR and fourth order gravity. While in GR a
spherically symmetric solution is, in any case, stationary and
static, here time-dependent evolution can be achieved depending on
the order of perturbations.

Hypothesizing a nonlinear Lagrangian we obtained a gravitational
attraction stronger than in GR. The hypothesis of dark matter is
needed in GR to have more gravitational attraction. Here without
hypothesis of alternative matter and without modifying the
gravitational constant we have qualitatively the same outcome.
This occurrence could be particularly useful to solve the problem
of missing matter in large astrophysical systems like galaxies and
clusters of galaxies. In fact dark matter could be nothing else
but the effects that GR, experimentally tested only up to Solar
System scales, does not work at extragalactic scales and then it
has to be corrected.

Then it is worth pointing out that $f(R)$-gravity seem good
candidate to solve several shortcomings of Modern Astrophysics and
Cosmology. Taking into account the results presented, it is clear
that only GR presents directly the Newtonian potential in the weak
field limit while corrections appear as soon as the theory is
non-linear in the Ricci scalar.

In forthcoming researches, there is the intention to confront such
solutions with experimental data in order to see if large
self-gravitating systems could be modelled by them and if the
experimental test of GR in the Solar System are compatible with
$f(R)$-gravity.

\section{Acknowledges}

The author would like to thank prof. Salvatore Capozziello for his
useful discussions and suggestions.


\begin{thebibliography}{99}

\bibitem{Weyl:1918}
         H. Weyl, Math. Zeit. \textbf{2}, 384  (1918)

\bibitem{Pauli:1919}
         W. Pauli, Phys. Zeit. \textbf{20}, 457 (1919)

\bibitem{Bach:1921}
         R. Bach, Math. Zeit. \textbf{9}, 110 (1921)

\bibitem{Weyl:1921}
         H. Weyl, Raum-Zeit-Materie, Springer Berlin (1921)

\bibitem{Eddington:1924}
         A.S. Eddington, \emph{The mathematical theory of
         relativity}, Cambridge University Press London (1924)

\bibitem{Lanczos:1931}
         C. Lanczos, Z. Phys. \textbf{73}, 147 (1931)

\bibitem{Buchdahl:1962}
         H.A. Buchdahl, Nuovo Cimento \textbf{23}, 141 (1962)

\bibitem{DeWitt:1965}
         B.S. de Witt, \emph{Dynamical theory of groups and fields},
         Gordon and Breach, New York (1965)

\bibitem{Bicknell:1974}
         G.V. Bicknell, Journ. phys. A \textbf{7}, 1061 (1974)

\bibitem{Havas:1977}
         P. Havas, Gen. Rel. Grav. \textbf{8}, 631 (1977)

\bibitem{Stelle:1978}
         K.S. Stelle, Gen. Rel. Grav. \textbf{9}, 353 (1977)

\bibitem{sneIa}
         A.G. Riess {\it et al.} Astron. J. {\bf 116}, 1009 (1998)\\
         A.G. Riess {\it et al.} Astroph. Journ. {\bf 607}, 665 (2004)\\
         S. Perlmutter {\it et al.} Astrophys. J. {\bf 517}, 565 (1999)\\
         S. Perlmutter {\it et al.} Astron. Astrophys. {\bf 447}, 31 (2006)

\bibitem{lss}
         S. Cole {\it et al.}, Mon. Not. Roy. Astron. Soc. {\bf 362},
         505 (2005)

\bibitem{cmbr}
         D.N. Spergel {\it et al.} Astrophys. J. Suppl. {\bf 148}, 175 (2003)\\
         D.N. Spergel {\it et al.} arXiv: astro-ph/0603449

\bibitem{GRGreview}
         S. Capozziello, M. Francaviglia, 0706.1146 [astro\,-\,ph] (2007)

\bibitem{OdintsovLadek}
         S. Nojiri, S.D. Odintsov, Int. J. Meth. Mod. Phys. {\bf 4},
         115 (2007)

\bibitem{farhoudi}M. Farhoudi, Gen. Relativ. Grav. \textbf{38},
         1261 (2006)

\bibitem{anderson}
         J.D. Anderson \emph{et al.} Phys. Rev. Lett. {\bf 81}, 2858 (1998)\\
         J.D. Anderson \emph{et al.} Phys. Rev. D {\bf 65}, 082004 (2002)

\bibitem{bertolami}
         O. Bertolami et al. arXiv: 0704.1733 [gr\,-\,qc] (2007)

\bibitem{Francaviglia}
         G. Magnano, M. Ferraris, M. Francaviglia, Gen. Rel. Grav. \textbf{19}, 465 (1987)\\
         G. Allemandi, A. Borowiec, M. Francaviglia, Phys. Rev. D \textbf{70}, 103503 (2004)\\
         M. Amarzguioui, O. Elgaroy, D.F. Mota, T. Multamaki, Astron.
         and Astrophys \textbf{454}, 707 (2006)

\bibitem{olmo}
         G. J. Olmo, Phys. Rev. D \textbf{72}, 083505, (2005)\\
         G. J. Olmo, Phys. Rev. Lett. \textbf{95}, 261102 (2005)\\
         G. J. Olmo, Phys. Rev. D \textbf{75}, 023511 (2007)

\bibitem{Damour:Esposito-Farese:1992}
         T. Damour, G. Esposito-Far\`{e}se, Class. Quantum Grav. \textbf{9}, 2093, (1992)

\bibitem{clifton}
         T. Clifton, Phys. Rev. D \textbf{77}, 024041 (2008)

\bibitem{landau}
         Landau Lev D., Lif\v{s}its E. M, \emph{Theoretical physics} vol. II

\bibitem{newtonian_limit_fR}
         S. Capozziello, A. Stabile, A. Troisi, Phys. Rev. D {\bf 76}, 104019 (2007)\\
         S. Capozziello, A. Stabile, A. Troisi, Modern physics letters A {\bf 24}, 659 (2009)

\bibitem{bransdicke}
         C. Brans and R.H. Dicke, Phys.\ Rev. {\bf 124}, 925 (1961)

\bibitem{spher_symm_fR}
         S. Capozziello, A. Stabile, A. Troisi, Class. Quant. Grav. {\bf 25}, 085004 (2008)

\bibitem{phd}
         A. Stabile, \emph{The Weak Field Limit of Higher Order Gravity}, PhD thesis, arXiv:0809.3570

\bibitem{newtonian_limit_R_Ric}
         S. Capozziello, A. Stabile, Class. Quant. Grav. {\bf 26}, 085019 (2009)

\bibitem{weinberg}
         S. Weinberg, \emph{Gravitation and Cosmology}, Wiley 1972, New York

\end{thebibliography}
\end{document}